\documentclass[12pt,notoc]{JHEP3}

\usepackage{amsmath,amssymb,euscript,array,cite,mathrsfs}

\usepackage{epsfig}

\def\XXint#1#2#3{{\setbox0=\hbox{$#1{#2#3}{\int}$}
     \vcenter{\hbox{$#2#3$}}\kern-.5\wd0}}

\def\a{\alpha}

\def\d{\delta}
\def\e{\epsilon}

\def\l{\lambda}
\def\m{\mu}
\def\n{\nu}

\def\r{\rho}
\def\s{\sigma}

\def\w{\omega}

\def\det{{\rm det}}

\def\BcalF{\boldsymbol{\cal F}}

\def\Bone{\boldsymbol{1}}

\def\BDelta{\boldsymbol{\Delta}}

\def\Bn{\boldsymbol{n}}

\def\Dbarslash{\,\,{\raise.15ex\hbox{/}\mkern-12mu {\bar D}}}
\def\Dslash{\,\,{\raise.15ex\hbox{/}\mkern-12mu D}}
\def\delslash{\,\,{\raise.15ex\hbox{/}\mkern-9mu \partial}}
\def\delbarslash{\,\,{\raise.15ex\hbox{/}\mkern-9mu {\bar\partial}}}

\def\rta{\rightarrow}
\def\orta{\overrightarrow}

\newcommand{\EQ}[1]{\begin{equation}\begin{split} #1 \end{split}\end{equation}}

\newcommand{\SP}[1]{\begin{equation}\begin{split} #1
\end{split}\end{equation}}

\title{`Superluminal' Photon Propagation in QED in Curved Spacetime is
Dispersive and Causal}
\author{Timothy J. Hollowood and Graham M. Shore\\
Department of Physics,\\ University of Wales Swansea,\\
Swansea, SA2 8PP, UK.\\
E-mail: {\tt t.hollowood@swansea.ac.uk, g.m.shore@swansea.ac.uk}}
\abstract{It is now well-known that vacuum polarisation
in QED can lead to superluminal low-frequency phase velocities 
for photons propagating in curved spacetimes. In a series of
papers, we have shown that this quantum phenomenon is dispersive
and have calculated the full frequency dependence of the refractive
index, explaining in detail how causality is preserved and various
familiar results from quantum field theory such as the Kramers-Kronig
dispersion relation and the optical theorem are realised in curved
spacetime.  These results have been criticised in a recent paper
by Akhoury and Dolgov \cite{Akhoury:2010hi}, who assert that
photon propagation is neither dispersive nor necessarily causal. 
In this note, we point out a series of errors in their work which 
have led to this false conclusion.
}

\begin{document}

\noindent {\bf 1.}~~ One of the important simplifications in our analysis 
of photon propagation in curved spacetime is the insight that, in the eikonal 
and `weak curvature' approximations (see section 2), the background may be 
replaced by its Penrose plane wave limit around the null geodesic describing 
the classical trajectory. The first main claim of \cite{Akhoury:2010hi}
is that such plane waves are too simple to manifest the vacuum 
polarisation induced modifications to photon propagation 
discovered by Drummond and Hathrell \cite{Drummond:1979pp}. This claim 
is simply not true: in fact plane waves have precisely the data that 
is encoded in the Drummond-Hathrell result.

The origin of superluminal low-frequency propagation is the 
effective action for QED in curved spacetime \cite{Drummond:1979pp}: 
\EQ{
{\cal L}=-\frac14F_{\mu\nu}F^{\mu\nu}+\frac{\alpha}{\pi} {1\over m^2}\Big(
a_1 R F_{\mu\nu}F^{\mu\nu} + a_2 R_{\mu\nu}
F^{\mu\lambda}F^\nu{}_\lambda+ a_3 R_{\mu\nu\lambda\rho}F^{\mu\nu}
F^{\lambda\rho}\Big)+\cdots\ .
\label{aa}
}
The dots indicate that this effective action is the first term in a 
derivative expansion, so results deduced from it are valid only for
low-frequency propagation.  The refractive index derived from \eqref{aa}
for a photon with wave-vector $k^\mu = \w\hat k^\m$, where $\w$ is the
frequency, is
\EQ{
n_{ij}=\delta_{ij}+
\frac{\alpha}{\pi} {1\over m^2}\big(c_1\delta_{ij} R_{\mu\nu} \hat k^\mu
\hat k^\nu+c_2R_{\mu i \nu j}\hat k^\mu \hat k^\nu\Big)\ ,
\label{ab}
}
where the indices $i,j$ label the two spacelike
polarisation directions. The constant coefficients $c_1,c_2$ are simply 
related to the known coefficients $a_2,a_3$ in the effective action.
If we now introduce a null coordinate $u$ along the direction $k^\mu$, then
\EQ{
n_{ij}=\delta_{ij}+
\frac{\alpha}{\pi} {1\over m^2}\big(c_1\delta_{ij} R_{uu}
+c_2R_{uiuj}\Big)\ .
\label{ac}
}
So the Drummond-Hathrell result depends on the curvature components
$R_{uiuj}$ and $R_{uu}$.

Now, a plane wave metric in Brinkmann coordinates takes the form
\EQ{
ds^2=2du\,dv+h_{ij}(u)y^i\,y^j\,du^2+dy^i\,dy^i\ .
\label{ad}
}
where $(u,v)$ are null coordinates and $y^i$ are transverse
spacelike coordinates.
The non-vanishing components of the Riemann tensor describe the wave 
profile, $R_{uiuj} = h_{ij}$. So if we apply the Drummond-Hathrell formula 
\eqref{ab} to a wave propagating along the geodesic $u$ with $v=y^i=0$, 
that is $k^u=\omega$, $k^v=0$ and $k^i=0$, we find a non-vanishing effect involving precisely the curvature components $R_{uiuj}$ that are non-trivial 
in a plane wave spacetime. 

Of course, this is apparent in our work 
\cite{Hollowood:2007kt, Hollowood:2007ku, Hollowood:2008kq, Hollowood:2009qz},
since we have demonstrated that the low-frequency limit of our full,
dispersive refractive index formula reproduces the original Drummond-Hathrell
result.  The error in \cite{Akhoury:2010hi} is in their eq.(12), 
which is derived simply from the equation of motion in a Ricci
flat plane wave background. This should read:
\EQ{
2 \partial_u \partial_v A_k ~+~ 
\bigl(g^{vv} - 8\hat a_3 R_{ukuk}\bigr) \partial_v^2 A_k ~=~0 \ ,
\label{ae}
}
with $\hat a_3 = {\a\over\pi}{1\over m^2}a_3$,
leading (since of course $g^{vv} = g_{uu} = 0$) to the usual result \eqref{ac}, and {\it not}, as quoted in \cite{Akhoury:2010hi},
\EQ{
\bigl(2 + 4\hat a_3 R_{ukuk}\bigr) \partial_u \partial_v A_k ~+~ 
g^{vv}\partial_v^2 A_k ~=~0 \ ,
\label{af}
}
The error is just a simple mistake of raising/lowering indices with the
off-diagonal metric in light cone coordinates $u,v$. However, it leads
the authors of \cite{Akhoury:2010hi} to the important but manifestly wrong conclusion that the Penrose limit fails to capture the Drummond-Hathrell effect.

The intuitive reason why the Penrose limit captures the essential geometry 
is most evident in the worldline, or proper-time, representations of the
propagator where it is clear that the vacuum polarisation depends 
on the geometry of geodesic fluctuations around the original photon
trajectory \cite{Hollowood:2007ku, Hollowood:2008kq}. This is precisely 
what is captured by the Penrose limit and is encoded in the Van Vleck-Morette 
matrix, which plays a key role in our analysis.

More recently \cite{Hollowood:2010bd}, we have also made an independent 
check on the validity of the Penrose limit. 
In the case of a background spacetime 
$dS_3 \times {\bf R}$, we can evaluate the refractive index, or spectral
density, exactly. Then, taking the appropriate WKB limits (section 2), we
recover precisely the result we obtain by using the Penrose limit directly,
in this case a conformally-flat symmetric plane wave with $R_{uu} < 0$.
This gives a highly non-trivial confirmation that the Penrose limit does
indeed retain the essential geometry of the full background field relevant
for determining the vacuum polarisation corrections to photon propagation.

\noindent {\bf 2.}~~Photon propagation with vacuum polarisation in curved
spacetime is characterised by three length scales and it is important to be
clear about the regimes of curvature and frequency where our results are valid.
Let $1/\sqrt{\cal R}$ denote the generic curvature length scale (so 
${\cal R}$ can represent a curvature or derivatives of curvature),
$\w$ the photon frequency and $m$ the electron mass. We work in the `weak
curvature' limit $\sqrt{\cal R} \ll m$ (so the curvature scale is much greater
than the electron Compton wavelength) and in the usual geometric optics, or
`eikonal' approximation $\w \gg \sqrt{\cal R}$ (so the photon wavelength
is much less than the curvature scale).  This defines the WKB limit in which
the Penrose limit is a good approximation to the general background
spacetime.

An important parameter in our analysis is the remaining dimensionless variable
$\w \sqrt{\cal R}/m^2$ and we need to consider all values of this to 
calculate the full frequency dependence of the refractive index.  Causality
is related to the limit $n(\infty)$ since this is identified with the wavefront
velocity of the light wave. 
Since ref.\cite{Akhoury:2010hi} misrepresents the role of this parameter in our
analysis of the high frequency behaviour of $n(\w)$, and therefore wrongly
criticises the implications of our results for causality, we spell out 
the parameter dependence of our results here in explicit detail.

Our result for the refractive index, taking scalar QED for simplicity, is
(see, e.g.~eq.(4.42) of \cite{Hollowood:2008kq})
\EQ{
\Bn(u;\omega)
=\Bone+\frac{\alpha}{2\pi\omega}\int_0^1
d\xi\,\xi(1-\xi)\BcalF\Big(u;
\frac{m^2}{2\omega\xi(1-\xi)}\Big)\ ,
\label{ba}
}
with
\SP{
\BcalF(u;z)&=\int_0^{\infty-i\epsilon}\frac{dt}{t^2}\,ie^{-izt}\,
\left[\Bone
-\BDelta\big(u,u-t\big)\sqrt{\det\BDelta\big(u,u-t\big)}
\right]\ ,
\label{bb}
}
The curvature dependence is embedded in the Van Vleck-Morette matrix
$\BDelta(u,u-t)$ which is a function of $\sqrt{\cal R}t$. For small $t$,
it can be expanded in schematic form as
\EQ{
\BDelta\big(u,u-t\big)\to\Bone+\sum_{n=1}^\infty
{\cal R}^n(u)t^{2n}\ ,
\label{bc}
}
where to linear order, the relevant curvature component is $R_{uiuj}$.
(The vector notation takes account of the polarisation dependence,
$i,j = 1,2$, and will be dropped from now on for simplicity.)
A simple rescaling $t \rta t/z$, or alternatively $t \rta t/\sqrt{R}$,
shows immediately that eq.\eqref{ba} can be written in the (clearly equivalent)
forms:
\EQ{
n~~=~~ 1 ~+~ {\a\over\pi} {m^2\over\w^2} 
F\Big({\w\sqrt{\cal R}\over m^2}\Bigr)
~~~~=~~ 1 ~+~ {\a\over\pi} {\sqrt{\cal R}\over\w} 
G\Big({m^2\over\w\sqrt{\cal R}}\Bigr) \ ,
\label{bd}
}
where the function $F$ (or $G$) is calculated non-perturbatively. This allows
us to access both the low and high frequency limits of the refractive index.

At low frequencies, we can expand $F$ as a power series in small 
${\w\sqrt{\cal R}\over m^2}$ using the VVM expansion \eqref{bc}. This starts
with a term of ${\cal O}\bigl({\w\sqrt{\cal R}\over m^2}\bigr)^2$, giving
the refractive index in the schematic form
\EQ{
n~~=~~ 1 ~+~ {\a\over\pi} {{\cal R}\over m^2} \biggl( 1 ~+~
{\cal O}\Bigl({\w\sqrt{\cal R}\over m^2}\Bigr) ~\biggr) \ .
\label{be}
}
The first term, independent of frequency, reproduces the Drummond-Hathrell 
result. Of particular interest is the next term in the expansion which, if 
present, gives a contribution to the imaginary part of the refractive index
${\rm Im}~n(\omega)$ of the form $\omega \partial_u R/m^4$. These contributions
can be reproduced by an extension of the DH effective action, discussed below.

At high frequencies, we can use the second form of \eqref{bd} and expand
$G$ for small ${m^2\over\w\sqrt{\cal R}}$. Here, the first term is found to be
of ${\cal O}(1)$ and the expansion may contain logarithms as well as powers.
We therefore find 
\EQ{
n~~=~~ 1 ~+~ {\a\over\pi} {\sqrt{\cal R}\over\w} ~+~ \ldots \ ,
\label{bf}
}
with $n(\infty) = 1$ as expected in a causal theory.

All this is illustrated quite explicitly in a simple example discussed in 
\cite{Hollowood:2008kq}, the conformally flat symmetric plane wave, where we
have an exact analytic expression for $\BcalF(z)$ in \eqref{bb}. In this model,
the wave profile function is $h_{ij} = \s^2 \d_{ij}$, so $\sqrt{\cal R} = \s$.
We find:
\EQ{
{\cal F}(z) ~~=~~ \s \Bigl( {z\over\s} \psi\Bigl(1 + {z\over2\s}\Bigr) 
- {z\over\s} \log{z\over2\s} - 1 \Bigr) \ .
\label{bg}
}
For large $z \sim {m^2\over\w}$, i.e.~low frequency, the digamma function has 
the expansion 
\EQ{
\psi\Bigl(1 + {z\over2\s}\Bigr) ~~=~~ \log{z\over2\s} + {\s\over z} - 
\sum_{n=1}^\infty {B_{2n}\over 2n} \Bigl({2\s\over z}\Bigr)^{2n} \ ,
\label{bh}
}
where $B_{2n}$ are the Bernoulli numbers, and so
\EQ{
{\cal F}(z) ~~=~~ -{1\over3} {\s^2\over z} 
\Bigl( 1 + {\cal O}\Bigl({\s\over z}\Bigr)^2 ~\Bigr) \ .
\label{bi}
}
Substituting in \eqref{ba} gives the refractive index \cite{Hollowood:2008kq}
\EQ{
n(\w) ~~=~~ 1 ~-~ {\a\over90\pi} {\s^2\over m^2} \Bigl( 1 + 
{\cal O}\Bigl({\w^2\s^2\over m^4}\Bigr) ~\Bigr) \ .
\label{bj}
}
Notice that in this case the expansion has no term linear in $\w$
and indeed ${\rm Im}~n(\w) = 0$ in this model.

In the opposite limit of small $z$, i.e.~high frequency, the relevant expansion
of the digamma function is 
\EQ{
\psi\Bigl(1 + {z\over2\s}\Bigr) ~~=~~ 
\sum_{n=0}^\infty {\psi_n(1)\over\Gamma(n+1)}
\Bigl({z\over2\s}\Bigr)^n \ ,
\label{bk}
}
giving
\EQ{
{\cal F}(z) ~~=~~ \s \Bigl( -1 - {z\over\s}\Bigl(\log{z\over2\s} 
+ \gamma \Bigr) + {\cal O} \Bigl({z\over\s}\Bigr)^2 \Bigr) \ .
\label{bl}
}
The refractive index is therefore \cite{Hollowood:2008kq}
\EQ{
n(\w) ~~=~~ 1 ~-~ {\a\over12\pi} {\s\over\w} 
\Bigl( 1 + {\cal O}\Bigl({m^2\over\w\s}\Bigr) ~\Bigr) \ ,
\label{bm}
}
where the first correction also includes the logarithm from \eqref{bl}. This
example demonstrates precisely how both the low and high frequency limits of the 
refractive index are realised in terms of the basic parameters $\sqrt{\cal R}$,
$m$ and $\w$ and gives the full non-perturbative function of 
$\w\sqrt{\cal R}/m^2$ that interpolates between them. It explicitly refutes 
the claim of ref.\cite{Akhoury:2010hi} that the corrections to the low-frequency
value $n(0)$ vanish in the high-frequency limit leaving $n(\infty) \neq 1$
with the associated problems with causality.

\noindent{\bf 3.}~~In flat spacetime, an important role is played by the
Kramers-Kronig dispersion relation
\EQ{
n(\infty) ~~=~~ n(0) ~-~ {2\over\pi} \int_0^\infty {d\w\over\w} {\rm Im}~n(\w) 
~~~~~~~~~~~~({\rm flat}~{\rm spacetime}) \ .
\label{ca}
}
In this case, ${\rm Im}~n(\w) > 0$ by virtue of the optical theorem,
which relates it to the forward scattering cross-section, so 
$n(\infty) < n(0)$ or equivalently $v_{\rm ph}(\infty) > v_{\rm ph}(0)$. 
A superluminal low-frequency phase velocity in flat spacetime would therefore
imply the wavefront velocity $v_{\rm ph}(\infty)$ exceeds
$c$, violating causality.\footnote{
An interesting, but quite separate, question is whether causality would be
violated in a model in which the strong equivalence principle is violated
explicitly by tree-level curvature couplings and we can use a Drummond-Hathrell
type action directly to analyse signal propagation and causality.

In ref.\cite{Shore:2003jx}, it was pointed out that this is related to
`{\it stable causality}' and depends on the global nature of the background
spacetime. The question is whether the spacetime still admits a global
Killing vector which is timelike with respect to the extended light cones 
of the DH action. Ref.\cite{Akhoury:2010hi} disputes this, repeating the 
proposal of Dolgov and Novikov \cite{Dolgov:1998fp} that the superluminal DH
effect permits the construction of a `time machine' which can be realised,
for example, by two black holes in relative motion where each exhibits
superluminal propagation in its vicinity. However, it has already been
explained in \cite{Shore:2002in}, section 5, (see also \cite{Konstantinov:1998qm})
that this proposed time machine does {\it not} work, precisely because of
the fact that the DH effect is dependent on the local spacetime curvature
(a fact dismissed in \cite{Akhoury:2010hi} as a `trivial complication'),
which invalidates the argument of \cite{Dolgov:1998fp}.}

However, in curved spacetime  neither the Kramers-Kronig dispersion relation
nor the optical theorem remain true in their normal flat spacetime forms,
and the usual insights based on \eqref{ca} are simply wrong. Nevertheless,
ref.\cite{Akhoury:2010hi} continues to use \eqref{ca} with the inevitable 
mistaken conclusions. Moreover, a confusion is made 
(eq.(2) of ref.\cite{Akhoury:2010hi})
between purely geometric and vacuum polarisation effects on ${\rm Im}~n(\w)$.

To explain this second point, note that the eikonal approximation for the 
solution to the ${\cal O}(\a)$ corrected wave equation is 
(e.g.~\cite{Hollowood:2009qz})
\EQ{
A_\m(x) ~~=~~ {\cal A}(x) \e_\m(x) e^{-i\w \bigl(V - \int^u du 
(n(u;\w) - 1)\bigr)} \ ,
\label{cb}
}
where $V$ is a Rosen coordinate.
The amplitude ${\cal A}(x)$ satisfies the classical equation
$\partial_u \log {\cal A} = - \hat\theta$ along the null geodesic, where
$\hat\theta$ is the optical scalar in the Raychoudhuri equations
\cite{Hollowood:2009qz} which describes the expansion or contraction of the
null geodesic congruence describing the wave propagation.
The amplitude can therefore increase or decrease due to this classical
geometric effect, which should be clearly distinguished from the ${\cal O}(\a)$
change to the amplitude which would be induced by an imaginary part of 
$n(u;\w)$ in \eqref{cb}. This would correspond to genuine dispersion due
to particle creation if ${\rm Im}~n(\w) > 0$ or the more subtle photon
`undressing' effect described carefully in ref.\cite{Hollowood:2010bd} if
${\rm Im}~n(\w) < 0$. It is only this latter effect that is related to
vacuum polarisation, the Kramers-Kronig dispersion relation and the optical
theorem. The assertion in \cite{Akhoury:2010hi} that such effects 
are `clearly negligible' in this context completely misses the point,
since the whole phenomenon of vacuum polarisation induced corrections to
photon propagation takes place at this order.

The form of the Kramers-Kronig dispersion relation which remains valid in
curved spacetime is \cite{Hollowood:2008kq}
\EQ{
n(u;\infty) ~~=~~ n(u;0) ~-~ {1\over{i\pi}} {\cal P} 
\int_{-\infty}^\infty {d\w\over\w}~n(u,\w) \ ,
\label{cc}
}
This assumes only that $n(u;\w)$ is bounded at infinity and analytic in the
upper half $\w$-plane, as required by causality. In order to recover the
conventional form \eqref{ca}, we need to assume also that we have
translation invariance along the geodesic, so that $n(\w)$ is an even function
of $\w$, and that it satisfies {\it real analyticity}, $n(\w^*) = n(\w)^*$.
This is described in full detail in \cite{Hollowood:2008kq}. One of the key
discoveries of our work is that this second property does
not hold in general, due to a novel analytic structure of the refractive
index, or vacuum polarisation, in curved spacetime. This is due to additional
singularities and branch cuts (evident in \eqref{bg}) related to the existence 
of conjugate points on the null geodesics, i.e.~points which may be joined by 
a continuous family of geodesics infinitesimally close to the original one. 
(In the world-line formalism, these correspond to zero-modes in the
fluctuations around the classical null geodesic path.)

The novel analytic structure we have uncovered in curved spacetime may also
give rise to a non-perturbative imaginary part for the refractive index,
distinct from the $\w{\partial_u}R/m^4$ type terms described above.
These contributions to ${\rm Im}~n(\w)$ can arise even in spacetimes where 
the curvature is translation invariant along the photon's null geodesic, as e.g.~the Ricci-flat symmetric plane wave \cite{Hollowood:2008kq}.

\vfill\eject

\noindent{\bf 4.}~~These entirely non-perturbative effects cannot be seen 
in a local effective action of the Drummond-Hathrell type, even when it is extended to include higher derivatives to all orders 
\cite{Shore:2002gw,Shore:2002gn,Hollowood:2009qz}. 
However, this is sufficient to capture the perturbative imaginary
parts proportional to $\w{\partial_u}R/m^4$. Since this is also relevant for
the central argument of ref.\cite{Akhoury:2010hi} (see section 5), we
recall some details of the derivation here.  The generalised effective action
is:
\SP{
\Gamma ~=~ &\int d^4x \sqrt{-g}~ \biggl[ 
-{1\over4}ZF_{\m\n}F^{\m\n} ~+~ 
D_\m F^{\m\l}\orta{d_0}D_\n F^\n{}_\l \\
&+~{1\over m^2}\Bigl(\orta{a_0} R F_{\m\n} F^{\m\n}~ 
+~\orta{b_0} R_{\m\n} F^{\m\l}F^\n{}_{\l}~
+~\orta{c_0} R_{\m\n\l\r}F^{\m\n}F^{\l\r} \Bigr) \\
&+~{1\over m^4}
\Bigl(\orta{a_1} R D_\m F^{\m\l} D_\n F^\n{}_{\l}~  
+~\orta{b_1} R_{\m\n} D_\l F^{\l\m}D_\r F^{\r\n} \\
&+~\orta{b_2} R_{\m\n} D^\m F^{\l\r}D^\n F_{\l\r}
~+~\orta{b_3} R_{\m\n} D^\m D^\l F_{\l\r} F^{\r\n} 
~+~\orta{c_1} R_{\m\n\l\r} D_\s F^{\s\r}D^\l F^{\m\n} 
~\Bigr)\biggr] 
\label{da}
}
In this formula, the $\orta{a_n}$, $\orta{b_n}$, 
$\orta{c_n}$ are known `form factor' functions of three operators, 
{\it i.e.}
\EQ{
\orta{a_n} \equiv a_n\Bigl({D_{(1)}^2\over m^2}, 
{D_{(2)}^2\over m^2}, {D_{(3)}^2\over m^2}\Bigr) \ ,
\label{db}
}
where the first entry $D_{(1)}^2$ acts on the first following term
(the curvature), etc. 

The refractive index derived from \eqref{da} is \cite{Shore:2002gn}
\EQ{
n_{ij}(\w) ~=~ \d_{ij} ~+~ 
\delta_{ij}
{\a\over\pi}{1\over m^2}c_1\Bigl({2i\w {\hat k}\cdot D\over m^2}\Bigr)R_{uu}~+~
{\a\over\pi}{1\over m^2}c_2\Bigl({2i\w {\hat k}\cdot D\over m^2}\Bigr)R_{uiuj} \ ,
\label{dc}
}
where the constant coefficients $c_1$, $c_2$ of \eqref{ac} are replaced
by functions of the operator ${\hat k}\cdot D \sim \partial_u$, which
describes the variation of the curvature tensors along the classical
null geodesic. The linear terms in the expansion of $c_1$ and $c_2$
give the contributions to ${\rm Im}~n(\w)$ of 
${\cal O}({\w\sqrt{\cal R}\over m^2}{R\over m^2})$ 
discussed following \eqref{be}. We have checked that these agree with
those derived using our VVM-based method \cite{Hollowood:2009qz}.

To understand the origin of this expression, consider as an example the 
term
\EQ{
{1\over m^4}  \int d^4 x \sqrt{-g} R_{\m\n\l\r}F^{\m\n}D^2 F^{\l\r} 
\label{dd}
}
in the effective action, incorporating a single power of $D^2$ from
the form factor. The corresponding contribution to the equation of
motion for $D_\m F^{\m\n}$ is 
\EQ{
{1\over m^4}\biggl[
-2(D^2 R_\m{}^\n{}_{\l\r})D^\m F^{\l\r}
-4(D_\s R_\m{}^\n{}_{\l\r})D^\s D^\m F^{\l\r}
-4 R_\m{}^\n{}_{\l\r} D^2 D^\m F^{\l\r}\biggr] \ .
\label{de}
}
The key point is that only the second term survives at leading order
in the eikonal approximation, where it contributes to the $c_2$ term
in \eqref{dc}.  The first term is suppressed by a power of 
${\cal R}/m^2$ while the third is of ${\cal O}(k^2)$, which is
suppressed by a power of $\a$ since the photon is on-shell
before the radiative corrections. By contrast, the second term is
of only of relative order ${\cal O}({\w\sqrt{\cal R}\over m^2})$, 
the now familiar parameter.

\noindent{\bf 5.}~~The main constructive part of
ref.\cite{Akhoury:2010hi} is based on a 1994 paper by Khriplovich
\cite{Khriplovich:1994qj} which argues on the basis of a proposed
general form for the graviton-photon-photon vertex that all
corrections to the refractive index beyond the low-frequency
contribution $n(0)$ vanish. Although at first sight plausible, it is
apparent that this argument misses the contributions of the form
$k^\m D_\m R/m^4 \sim \w\partial_u R/m^4$ that we have found in both
the effective action and VVM approaches. This was pointed out
already in ref.\cite{Shore:2002gn}. Of course, Khriplovich's
method could not in any case detect the non-perturbative 
contributions following from the occurrence of conjugate points
on the photon's null geodesic.

Although it is obvious from our explicit construction of $n(\w)$ in
many examples that Khriplovich's argument cannot be correct, the 
loophole in the analysis of \cite{Khriplovich:1994qj} is in fact
quite subtle. Following \cite{Khriplovich:1994qj}, we define the 
graviton-photon-photon vertex in momentum space on a flat background:
\EQ{
(2\pi)^4 \d^{(4)}(q+k+k') \Gamma(q,k,k') ~~=~~\tilde h^{\m\n}(q)
\int dz~ e^{iz.q}\langle\e(k)|T_{\m\n}(z)|\e(k')\rangle \ ,
\label{ea}
}
where $\e(k)$ is the photon polarisation and $\tilde h_{\m\n}(q)$ 
is the external graviton field.
Then, according to \cite{Khriplovich:1994qj}
(see also \cite{Berends:1975ah,Drummond:1979pp}), 
the most general form 
of the vertex to ${\cal O}(1/m^2)$ can be written as 
\SP{
\Gamma(q,k,k') ~=~ &a_1(q^2,k^2,k^{\prime 2}) 
\tilde R^h(q) \tilde F_{\m\n}(k) \tilde F^{\m\n}(k') ~+~
a_2(q^2,k^2,k^{\prime 2}) 
\tilde R^h_{\m\n}(q) \tilde F^{\m\l}(k) \tilde F^{\n}{}_\l(k')\\ 
&+~ a_3(q^2,k^2,k^{\prime 2}) 
\tilde R^h_{\m\n\l\r}(q)\tilde F^{\m\n}(k) \tilde F^{\l\r}(k') \ ,
\label{eb}
}
where $\tilde F_{\m\n}(k) = k_\m \e_\n(k) - k_\n \e_\m(k)$
and $\tilde R^h(q)$ is the Fourier transform of 
the Ricci scalar evaluated with the metric identified with the graviton 
field, etc. The $a_i(q^2,k^2,k^{\prime 2})$ are form factors. The similarity
with the effective action \eqref{da} is clear. Higher-order Lorentz
structures of ${\cal O}(1/m^4)$ with four momenta can be read off from
\eqref{da} (see also, e.g.~\cite{Armillis:2009pq}).

The argument of \cite{Khriplovich:1994qj, Akhoury:2010hi} now proceeds 
by claiming that the imaginary part of $\Gamma$, with the photons taken
on-shell, only has a contribution from 
$q^2 = 0$ so that ${\rm Im}\Gamma(q^2,0,0) \sim \d(q^2)$. Since the vertex
has no dependence on the photon frequencies, it follows that the refractive
index itself must be independent of frequency and that $n(\w) = n(0)$
for all $\w$, i.e.~the Drummond-Hathrell effect is non-dispersive.

To see what is wrong with applying this argument in curved spacetime,  
notice that for a slowly varying background gravitational field, the 
Fourier transform $\tilde R(q)$ of the curvature is in fact very singular 
at $q=0$. If we expand (dropping indices for clarity):
\EQ{
R(z) ~~=~~R(0) ~+~ z^\m \partial_\m R(0) ~+~ \ldots \ ,
\label{ec}
}
then 
\EQ{
\tilde R(q)~~=~~ R(0)\d^{(4)}(q)~-~
i\partial_\m R(0) {\partial\over\partial q_\m} \d^{(4)}(q)+\cdots\ .
\label{ed}
}
To see the effect of this singular behaviour involving derivatives
of the delta function, substitute \eqref{ed} and integrate a typical 
vertex term over $q$:
\SP{
&\int d^4 q~ a(q^2, k^2, (k+q)^2) ~\tilde R(q) \tilde F(k) \tilde F(-k-q)\\
&=~~a(0,k^2,k^2)~ R(0) \tilde F(k) \tilde F(-k) ~-~
a^{(3)}(0,k^2,k^2)~ 2i k^\m\partial_\m R(0) ~\tilde F(k) \tilde F(-k) \\
&~~~~+~a(0,k^2,k^2) ~i\partial_\m R(0) ~\tilde F(k)
{\partial\over\partial k_\m}\tilde F(-k) \ ,
\label{ee}
}
where $a^{(3)}$ denotes differentiation w.r.t.~the third argument.
Of these terms, the first is obviously just the zero-momentum
Drummond-Hathrell contribution, while the third is lower order in the
eikonal approximation. The second term, however, can contribute
at leading order even on-shell and gives rise to an imaginary
part for the refractive index proportional to $k^\m\partial_\m R(0)$,
i.e.~precisely the frequency-dependent contribution of 
$\w \partial_u R/m^4$ type that we have already identified. 
It arises in essentially the same way as illustrated for the 
effective action in \eqref{dd},\eqref{de}.

What this shows is that when the Khriplovich argument is applied to 
curved spacetime backgrounds, it is necessary to keep careful track
of the full singularity structure, including derivatives of the
momentum-space delta function associated with the graviton insertion.
Yet again, we find that a careful analysis shows that
the refractive index for photon propagation on curved spacetime
{\it is} dispersive, with a low-frequency expansion \eqref{be} which 
may have an imaginary part proportional to $k^\m D_\m R/m^4$.

\noindent{\bf 6.}~~Finally, the question of whether QED in curved spacetime
is causal is ultimately determined by whether the commutator, or Pauli-Jordan,
Green function $iG_{\m\n}(x,x') = \langle 0|[A_\m(x),A_\n(x')]|0\rangle$
vanishes outside the light cone. This was demonstrated in 
ref.\cite{Hollowood:2008kq} by an explicit construction, contradicting the
claims in \cite{Akhoury:2010hi}. In fact, we were able to show that the 
commutator in scalar QED in the Penrose plane wave limit, incorporating vacuum polarisation, can be written as
\SP{
&G_{ij}(x,x')~~=~~2\frac{\alpha}{\pi}\,
\int_{u'}^u d\tilde u\,\int_0^{u-u'}
\frac{dt}{t^2}\,\int_0^1d\xi\, \xi(1-\xi)\,\\ &~~~~
\times
\Delta_{ij}(\tilde u,\tilde u-t)
\sqrt{\det\BDelta(\tilde u,\tilde u-t)} ~ 
G\left(\frac{m^2t}{2\xi(1-\xi)(u-u')};x,x'\right)+\cdots \ ,
\label{fa}
}
whee the omitted terms are independent of the curvature.
Here, $G\left(m^2;x,x'\right)$ is the
commutator Green function for a free massive scalar field, which clearly has
support only on or inside the light cone. Since $G_{ij}(x,x')$ therefore
vanishes outside the light cone, causality is manifest.

It is interesting to relate this to the refractive index. 
If we take $x = (u,V,0,0)$ and $x'= (u'\rta -\infty,0,0,0)$ to be
two points on the classical photon trajectory, we find (note that
since we are taking $u>u'$, the commutator and retarded Green functions 
coincide):
\EQ{
G_{ij}(x;x')~\thicksim~
\int_{-\infty}^u d\tilde u\,\int_{-\infty}^{\infty}
  d\omega \,n_{ij}(\tilde u;\omega)e^{-i\omega V}\ .
\label{fb}
}
The properties we have established for the refractive index, in particular
analyticity in the upper half $\w$-plane, now show that the commutator 
(retarded) Green function vanishes when $V<0$, i.e.~outside the light cone.

\vskip0.3cm
In summary, despite the many erroneous claims to the contrary in 
ref.\cite{Akhoury:2010hi}, we have clearly established  
\cite{Hollowood:2007kt, Hollowood:2007ku, Hollowood:2008kq, Hollowood:2009qz}
that photon propagation in QED in curved spacetime is indeed dispersive 
and causal.

\end{document}